\documentclass[twocolumn,superscriptaddress,amsmath,amssymb,
floatfix,aps,prl,showpacs]{revtex4}
\usepackage{graphics,graphicx}

\usepackage[usenames]{color}
\definecolor{BrickRed}{cmyk}{0,0.89,0.94,0.28}
\definecolor{MidnightBlue}{cmyk}{0.98,0.13,0,0.43}
\definecolor{DarkGreen}{rgb}{0,0.7,0.1}
\newcommand{\add}[1]{{#1}}\newcommand{\comm}[1]{{}}
\newcommand{\addAK}[1]{{#1}}
\newcommand{\addTwo}[1]{{#1}}

\begin{document}

\title{Thymic selection of T-cell receptors as an extreme value problem}

\author{Andrej Ko\v smrlj}
\affiliation{Department of Physics, Massachusetts Institute of Technology,
Cambridge, MA 02139, USA}

\author{Arup K. Chakraborty}
\affiliation{Departments of Chemical Engineering, Chemistry and
Biological Engineering, Massachusetts Institute of Technology,
Cambridge, MA 02139, USA}

\author{Mehran Kardar}
\affiliation{Department of Physics, Massachusetts Institute of Technology,
Cambridge, MA 02139, USA}

\author{Eugene I. Shakhnovich}
\affiliation{Department of Chemistry and Chemical Biology, Harvard
University, Cambridge, MA 02138, USA}

\date{\today}

\begin{abstract}

T lymphocytes (T cells) orchestrate adaptive immune responses upon
activation. T cell activation requires sufficiently strong binding of T cell
receptors (TCRs) on their surface to short peptides (p) derived from
foreign proteins, which are bound to major histocompatibility (MHC)
gene products (displayed on antigen presenting cells). A diverse and
self-tolerant T cell repertoire is selected in the thymus. We map thymic
selection processes to an extreme value problem and provide an analytic
expression for the amino acid compositions of selected TCRs (which
enable its recognition functions).

\end{abstract}

\maketitle
The adaptive immune system clears pathogens from infected hosts
with the aid of T lymphocytes (T cells). Foreign (antigenic) and
self-proteins are processed into short peptides (p) inside
antigen-presenting cells (APC), bound to MHC proteins, and
presented on the surface of APCs. Each \add{T-cell receptor (TCR)}
has a conserved region participating in the signaling functions, and a
\add{highly variable segment} 
responsible for antigen recognition. Because variable regions are
generated by stochastic rearrangement of the relevant genes, most
T cells express a distinct TCR. The diversity of the T cell repertoire
enables the immune system to recognize many different antigenic
short pMHC complexes. Peptides presented on MHC class I  are
typically 8--11 amino acids long~\cite{kuby}, which is enough to
cover all possible self-peptides (the human proteome consists of
$P\approx 10^7$ amino-acids~\cite{borroughs04, flicek08}) as
well as many antigenic peptides. TCR recognition of pMHC is
both specific and degenerate. It is specific, because most
mutations to the recognized peptide amino acids abrogate
recognition~\cite{huseby06,huseby05}. It is degenerate
because a given TCR can recognize several antigenic
peptides~\cite{unanue84}.

The gene rearrangement process ensuring the diversity of TCR
is random. It may thus result in T cells potentially harmful to the
host, because they bind strongly to \add{self peptide-MHC
complexes}; 
or useless T cells which bind too weakly to MHC
to recognize antigenic peptides. Such aberrant TCRs
are eliminated in the thymus~\cite{boehmer03,
palmer03,
siggs06,
hogquist05},
where immature T cells (thymocytes) are
exposed to a large set  ($10^3-10^4$) of self-pMHC. Thymocytes
expressing a TCR that binds with high affinity to any self-pMHC
molecule are deleted in the thymus (a process called negative
selection). However, a thymocyte's TCR must also bind sufficiently
strongly to at least one self pMHC complex to receive survival
signals and emerge from the thymus (a process called positive
selection).

Signaling events, gene transcription programs, and cell migration
during T cell development in the thymus ~\cite{boehmer03,
palmer03,
siggs06,hogquist05,
daniels06, bousso02,
borghans03, scherer04, detours99tb, detours99pnas} have been
studied extensively. Despite many important advances, how
interactions with self-pMHC complexes in the thymus shape the
peptide-binding properties of selected TCR amino acid sequences,
such that mature T cells exhibit their special properties, is poorly
understood. To address this issue, \add{in Ref.~\cite{kosmrlj08}}
we \add{numerically studied a simple model} where TCRs
and pMHC \add{were represented} by strings of amino acids
(Fig.~\ref{fig:model}). These strings indicate the amino-acids on
the interface between TCRs and pMHC complexes, and it is
assumed that each site on a TCR 
interacts only with a corresponding site on pMHC. The binding
interface of TCR is \add{actually} composed of a region that is in
contact with \add{the} MHC \add{molecule}, and a segment that is
in contact with the peptide. \add{It is the latter part that is highly
variable, while the former is more conserved.} We shall therefore
explicitly consider \add{only} the former amino-acids, but not the latter.
Similarly, there are many possible peptides that can bind to
MHC, and their sequences are considered explicitly, whereas
those of the MHC are not (there are only a few types of MHC in
each individual human~\cite{kuby}). We could in principal add
a few sites to the TCR and pMHC strings to account for 
any variability in the segments not considered.

Simplified representations of amino-acids (e.g., as a string of
numbers or bits) were employed earlier~\cite{detours99tb,
detours99pnas, chao05} in the context of TCR-pMHC interactions,
mainly to \addAK{report that} negative selection reduces TCR
cross-reactivity. In \add{Ref.}~\cite{kosmrlj08}, we
\add{{\em numerically} studied} the model in Fig.~\ref{fig:model}
(and described below) to qualitatively
describe the role of positive and negative selection on the
amino-acid composition of selected TCRs. By randomly generating
TCR and pMHC sequences, and implementing thymic
selection {\add{ in silico}, we showed that selected TCRs  are enriched
in weakly interacting amino acids, and explained how this leads
to specific, yet cross-reactive, TCR recognition of
antigen\addAK{, a long-standing puzzle}. 
In this paper we \add{show that the model can be solved
{\em exactly} in the  limit of long TCR/peptide sequences. 
 The resulting analytic expression for  the amino-acid composition
 of selected TCRs  is surprisingly accurate even for short peptides and 
 \addAK{provides a theoretical basis for} previous numerical results.
Furthermore, we are able to obtain a phase diagram that indicates
the ranges of parameters where negative or positive selection are
dominant, leading to quite different bias  in selection/function.}
\begin{figure}[t]
\includegraphics[width=7.6cm]{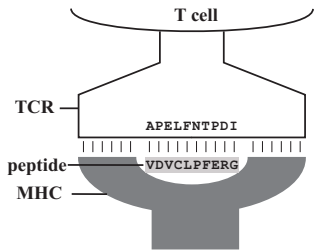}
\caption{
Schematic representation of the interface between TCR and
pMHC complexes. The segment of TCR that is in contact with
peptides is highly variable and modeled by a string of $N$
amino-acids. The peptide is also modeled by a sequence of
length $N$, and the binding energy is computed as a sum of
pairwise interactions. We don't explicitly consider TCR sites
in contact with MHC, as they are more or less
conserved, and only assign them a net interaction energy
$E_c$.} 
\label{fig:model}
\end{figure}

To assess the effects of thymic selection, as well as antigen
recognition, we evaluate the free energy of interaction between
TCR-pMHC pairs (for brevity, free energy will be referred to as
energy). The interaction energy is composed of
two parts: a TCR interaction with MHC, and a TCR interaction
with the peptide. The former is given a value $E_c$ (which
may be varied to describe different TCRs and MHCs). The
latter is obtained by aligning the TCR and pMHC amino-acids
that are treated explicitly, and adding the pairwise interactions
between corresponding pairs. For a given TCR-pMHC pair,
this gives
\begin{equation}
E_{\mathrm{int}}\bigl(\vec t, \vec s \ \!\bigr) = E_c
+ \sum_{i=1}^N J(t_i,s_i),
\label{eq:int}
\end{equation}
where $J(t_i,s_i)$ is the \add{contribution from} the
$i$th amino acids of the TCR ($t_i$) and the peptide ($s_i$),
and $N$ is the length of the variable TCR/peptide region.
The matrix $J$ encodes the interaction energies between
specific pairs of amino-acids. For numerical implementations
we use the Miyazawa-Jernigan (MJ) matrix~\cite{miyazawa96}
that was developed in the context of protein folding.

Immature T cells interact with a set ${\cal S}$ of $M$ self-pMHC
\add{complexes}, where typically $M$ is of the order of $10^3-10^4$.
\add{To mimic thymic selection, sequence\addAK{s}} that bind to
any self-pMHC too strongly
($E_{\mathrm{int}} < E_n$) are deleted (negative selection).
However, a thymocyte's TCR must also bind sufficiently
strongly ($E_{\mathrm{int}} < E_p$) to at least one
self-pMHC to receive survival signals and emerge from the
thymus (positive selection).
A thymocyte expressing TCR with string $\vec t$ will
thus be selected if {\em  the strongest interaction}
with self-pMHC is between thresholds for negative and positive
selection, i.e.
\begin{equation}
E_n  < \min_{\vec s \in \cal S} \bigl\{E_{\mathrm{int}}
\left(\vec t, \vec s \ \!\right) \bigr\} < E_p.
\label{eq:selection}
\end{equation}
Recent experiments~\cite{daniels06} show that the difference
between thresholds for positive and negative selection is
relatively small (a few $k_B T$). 

Equation~(\ref{eq:selection})
\add{casts} thymic selection as an extreme
value problem~\cite{EVD},
\add{enabling us} to calculate the probability
$P_\mathrm{sel} ({\vec t }\,)$ that a TCR sequence $\vec t$
will be selected in the thymus.
\add{Let us indicate by $\rho(x|\vec t \,)$ the probability density
function (PDF) of the interaction energy between the TCR $\vec t$
and a random peptide. The PDF  $\Pi (x|{\vec t }\,)$ of the strongest
(\add{minimum)} 
of the $M$ independent random interaction energies
is then obtained by multiplying $\rho$ with the probability of all
remaining $(M-1)$ energy values being
larger-- $ \left (1 - P\!\left(E < x|\vec t \ \!\right) \right)^{M-1}$,
where $P(E < x|\vec t \,)$ is the cumulative probability-- and noting
the multiplicity $M$ for which energy is the lowest.}
The probability that TCR $\vec t$ is selected is then obtained by
integrating $\Pi (x|{\vec t }\,)$ over the allowed range, as
\begin{eqnarray}
P_\mathrm{sel} \!\left(\vec t \ \!\right)& = & 
\int_{E_n}^{E_p} \Pi\!\left(x|\vec t \ \!\right) dx,
\quad{\rm with}\nonumber \\
\Pi \!\left(x|\vec t \ \!\right) & = &
M\ \rho\!\left(x|\vec t \ \!\right) \left (1 -
P\!\left(E < x|\vec t \ \!\right) \right)^{M-1}.
\end{eqnarray}
\add{For $M\gg 1$, this extreme value distribution (EVD) converges
to one of three possible forms,~\cite{EVD}
depending on the tail of the PDF for each entry. 
Equation~(\ref{eq:int}) indicates that in our case as each energy is
the sum of $N$ 
contributions, $\rho(x|\vec t \,)$ should be a Gaussian for large $N$,
in which case the relevant EVD is the Gumbel distribution.~\cite{EVD}}

\add{To obtain an explicit form for $\Pi (x|{\vec t }\,)$, we model
the set $\cal S$ of self-peptides as $M$ strings in which each 
amino-acid is chosen independently. The probability $f_a$ for
selecting amino-acid $a$ at each site is taken to be the frequency
of this amino-acid in the self-proteome.
For a specific TCR sequence $\vec t$, the average interaction energy
with self peptides follows from Eq.~(\ref{eq:int}) as
$E_{\mathrm{av}} ({\vec t }\,) = E_c + \sum_{i=1}^N \mathcal{E}  (t_i)$,
with $\mathcal{E}(t_i)  = \left[  J(t_i,a)\right]_a$,
where we have denoted the average over self amino-acid frequencies
by $\left[G(a)\right]_a\equiv \sum_{a=1}^{20} \addAK{f_a} G(a)$.
Similarly, the variance of the interaction energy is 
$V({\vec t }\,) = \sum_{i=1}^N \mathcal{V} (t_i)$,
where
$\mathcal{V} (t_i) = \left[ J(t_i,a)^2\right]_a-\left[ J(t_i,a)\right]_a^2$.
} 
\add{For large $N$, we can approximate}
$\rho(x|{\vec t }\,)$ with a Gaussian \add{PDF} with
\add{the above} mean and variance.
\add{From standard results for the Gumbel distribution~\cite{EVD}, we
conclude that in} the limit of $M \gg 1$,
the peak of the distribution $\Pi (x|{\vec t }\,)$ is located at
\begin{equation}
E_0 \!\left({\vec t }\,\right) = E_{\mathrm{av}}\! \left({\vec t }\,
\right) - \sqrt{2 V \!\left({\vec t }\,\right) \ln M}\,,
\label{eq:peak}
\end{equation}
and \add{its} width is
$\Sigma_0 (\vec t \,) = \sqrt{\pi^2 V(\vec t \,)/ (12 \ln M)}$.
(Since the PDF $\rho(x|\vec t \,)$ originates from a \add{bounded}
set of energies, it is strictly not Gaussian in the tails. Hence,
once the extreme values begin to probe the tail of the
distribution, the above results will no longer be valid. Indeed,
in the limit when $M \sim \mathcal{O} (20^N)$, the EVD will
approach a delta-function centered at the $M$--independent
value corresponding to the optimal
binding energy.)
 
In the limit of long TCR/peptides ($N \gg 1$), we can exactly
calculate the statistics of the amino-acid composition of
selected TCRs. To obtain a proper thermodynamic limit, we
need to set $\{E_c,E_p,E_n\}\propto N$, {\em and}
$\ln M\propto N$. The latter ensures that the peak of the
distribution, $E_0(\vec t \,)$, is proportional to $N$, and also 
results in a width $\Sigma_0(\vec t \,)$ which is independent
of $N$. (The relation $\ln M=\alpha N$ can be justified 
with the expectation that $M$ should grow proportionately to
the proteome size $P$, while $N\propto\ln P$ to enable
encoding the proteome.) In \add{this} large $N$ limit, the EVD is
sufficiently narrow that the value of the optimal energy can be
precisely equated with the peak $E_0 (\vec t \,)$, and
Eq.~(\ref{eq:selection}) for the selection condition can be
replaced with
\begin{equation}
E_n < E_c + \sum_{i=1}^N \mathcal{E} (t_i)
- \sqrt{2 \ln M \sum_{i=1}^N \mathcal{V}(t_i)} < E_p \,.
\label{eq:selection2}
\end{equation}
Thus, for each sequence $\vec t$, we have to evaluate the
`Hamiltonian' $E_0 (\vec t \,)$, and the sequence is accepted
if this energy falls in the interval $(E_n,E_p)$. This is somewhat
similar to the micro-canonical ensemble in Statistical Physics,
with the restriction of the energy to an interval rather than a
narrow range only a minor elaboration (see below). From the
equivalence of canonical and micro-canonical ensembles for
large $N$, we know that the probability for a sequence is
governed by the Boltzmann weight
\addTwo{$p(\vec t \,) \propto
\left( \prod_{i=1}^N f_{t_i} \right)
\exp[-\beta E_0 (\vec t \,)]$.
Here $\{f_a\}$ indicate the natural
frequencies of the different amino-acids prior to selection,
while the effect of thymic selection is
captured in the parameter $\beta$ which is determined by
solving for the average energy.}

The appearance of
$\sqrt{2\ln M \sum_i \mathcal{V}(t_i)}$ in the Hamiltonian
\add{initially appears as a complication
that makes exact computation of the average energy from
$\exp[-\beta E_0 (\vec t \,)]$ impossible}. However, this apparent
`coupling' is easily dealt with by standard methods such as Legendre
transforms or Hamiltonian minimization~\cite{kardar83}.
\add{This can be justified easily as follows: We need to solve a
`Hamiltonian' $\cal H(U,V)$ which depends on two extensive
quantities $U=\sum_{i=1}^N \mathcal{E} (t_i)$
and $V= \sum_{i=1}^N \mathcal{V}(t_i)$. 
The corresponding partition function can be decomposed as 
$Z=\sum_{U,V}\Omega(U,V)e^{-\beta{\cal H}(U,V)}$, but can be
approximated with its largest term. Note that the same density of
states  $\Omega(U,V)\equiv e^{S(U,V)/k_B}$ appears,
 irrespective of the specific form of ${\cal H}(U,V)$.  
In particular, the choice \addTwo{${\cal H}_0=E_c +
U-\gamma V - \ln M / (2 \gamma)=
E_c + \sum_{i=1}^N[(\mathcal {E} (t_i) - \gamma
\mathcal{V} (t_i)]
- \ln M / (2 \gamma) $}
corresponds to a set of non-interacting variables, with
\begin{equation}
p(\vec t \,) \propto  \prod_{i=1}^N f_{t_i}\exp \bigg[
- \beta \big(\mathcal {E} (t_i) -
\gamma \mathcal{V} (t_i) \big)\bigg],
\label{eq:probms}
\end{equation}
for which thermodynamic quantities (such as entropy) are
easily computed.
By judicious choice of $\gamma$ we can then ensure that the
same average energy appears for \addTwo{${\cal H}_0 (\vec t \,)$}
and our $E_0 (\vec t \,)$.
\addTwo{Using Legendre transforms, which
is equivalent to minimizing ${\cal H}_0(\vec t \,)$ with respect to
$\gamma$, one finds that the required $E_0(\vec t \,)$ is obtained
by setting}
\addTwo{$\gamma (\beta) = \sqrt{\ln M / (2 N
\left< \mathcal{V}\right>_{\beta, \gamma} )}$,}
where $\left< \cdots\right>_{\beta, \gamma}$
refers to the average with the non-interacting weight
$e^{-\beta (U-\gamma V)}$.
}

\add{Finally, the value of $\beta$ has to be determined by
constraining 
the average energy determined above to the range in
Eq.~(\ref{eq:selection2})\addAK{, while maximizing entropy.}}
Given the bounded set of
energies, the inverse temperature $\beta$ can be either
negative or positive. The $20^N$ possible values for
$E_0 (\vec t \,)$ span a range from $E_{\mathrm{min}}$ to
$E_{\mathrm{max}}$, and a corresponding number of states
which is a bell-shape between these extremes with a
maximum at some $E_{\mathrm{mid}}$. If
$E_{\mathrm{mid}} > E_p$, we must set $\beta$ such that 
$\left<E_0 (\vec t \,) \right>=E_p $. In this case, $\beta > 0$,
positive selection is dominant and stronger amino-acids are
selected. If $E_{\mathrm{mid}} < E_n$, we must set $\beta$
such that  $\left<E_0 (\vec t \,) \right>=E_n$, $\beta < 0$,
negative selection is dominant and weaker amino-acids are
selected. For $E_n < E_{\mathrm{mid}} < E_p$, we must set
$\beta =0 $ and there is no modification due to selection. 

\begin{figure}[b]
\includegraphics[width=8.6cm]{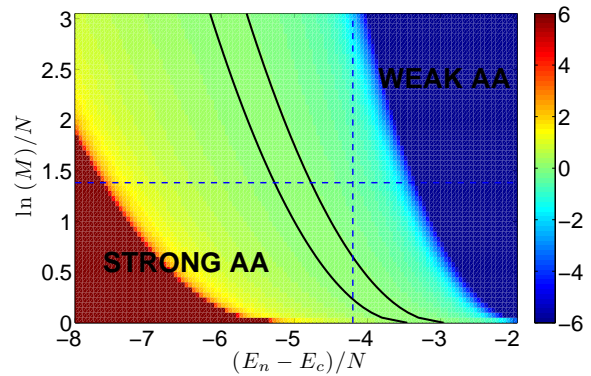}
\caption{
(Color) Color representation of the dependence of the
inverse temperature $\beta$ on the number of self-peptides
$\ln M/N$ and the threshold for negative selection energy
$E_n/N$ with $(E_p - E_n)/N=0.5 k_B T$ in the limit of large $N$.
The region between the black lines corresponds to $\beta=0$,
to the right (left) of which negative (positive) selection is dominant,
and weak (strong) amino-acids are selected. Note that 
as $(E_p-E_n)/N$ goes to zero, the intermediate region disappears.
The dotted lines
indicate the relevant parameter values for thymic selection
in mouse \addAK{(see text)} that result in $\beta = -0.37 (k_B T)^{-1}$.
}

\label{fig:phase_diagram}
\end{figure}
Figure~ \ref{fig:phase_diagram} depicts the variation of $\beta$
as a function of $\ln(M)/N$ and threshold
for negative selection $E_n$ with $(E_p - E_n)/N=0.5 k_B T$.
\addAK{Consider TCRs that do not bind too strongly
or weakly to MHC, as such TCRs are unlikely to be selected
(e.g., $E_n - E_c = -21 k_B T$).}
For the set
of parameters that are relevant for
thymic selection in mouse~\cite{kosmrlj08}, i.e. $N=5$,
$E_p - E_n = 2.5 k_B T$ and $M=10^3$,
we find $\beta = -0.37 (k_B T)^{-1}$ which means that negative
selection is dominant and weaker amino-acids are selected. Also
$\gamma=0.83 (k_B T)^{-1}$ indicating a preference for amino-acids
with smaller variations in binding energy. With these parameters we
can calculate the amino-acid frequencies of selected TCRs as
\begin{equation}
f_a^\mathrm{(sel)} = \frac{f_a \exp \big[-\beta \big(\mathcal{E} (a)
- \gamma \mathcal{V} (a) \big) \big] } 
{\sum_{b=1}^{20}f_b \exp \big[-\beta \big(\mathcal{E} (b)
- \gamma \mathcal{V} (b) \big) \big] }.
\label{eq:fsel}
\end{equation}
\add{It is important to ask if the above expression, exact in the limit
of $N\to\infty$, has any relevance to the actual TCR/peptides with
$N\sim 5-10$. We thus numerically simulated the case of $N=5$
by generating a random set of $10^6$  TCR sequences, and selected
them against $M=10^3$ self-peptides. The selected TCRs were used
to construct the amino-acid frequencies depicted in
Fig.~\ref{fig:composition}. The dashed line in this figure comes from
Eq.~(\ref{eq:fsel}) with the same $J$ and $\{f_a\}$. The agreement
between the two is remarkable given the small value of $N=5$, and
may be indicative of small corrections to the $N\to\infty$ result.
}
\begin{figure}[t]
\includegraphics[width=8.6cm]{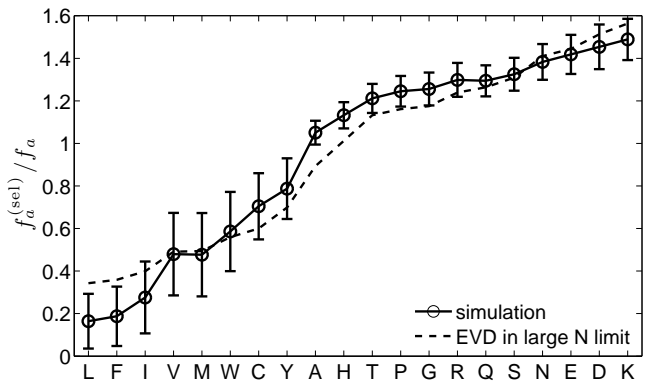}
\caption{
Amino-acid composition of selected TCR sequences, ordered in
increasing frequency along the abscissa. The data points in black
are obtained numerically with the parameters relevant to mouse
(see caption of Fig.~\ref{fig:phase_diagram}, and text). The error
bars reflect the sample size used to generate the histograms and
differences for different realizations of $M$ self-peptides. The
dashed line is the result of the EVD analysis in the large $N$ limit
from Eq.~(\ref{eq:fsel}), and the agreement is quite good. In both
cases we have used the Miyazawa-Jernigan matrix
$J$~\cite{miyazawa96}, and amino acid frequencies $f_a$ from
the mouse proteome~\cite{flicek08,kosmrlj08}.
}
\label{fig:composition}
\end{figure}

\add{
Equation~(\ref{eq:fsel}) thus provides an analytical expression that
captures the characteristics of TCR amino-acids selected against
many peptides in the thymus. In accord with previous numerical
results~\cite{kosmrlj08}, and some available data from normal
mouse, and human, it predicts (since $\beta<0$) that TCR sequences
are enriched in weakly interacting amino-acids (small $\mathcal{E}$).
This result was used previously~\cite{kosmrlj08} to explain their specificity.
However, Eq.~(\ref{eq:fsel}) further indicates the role of promiscuity 
of amino-acids (captured by the parameter $\gamma$) which was
not elucidated from the limited numerical data. Furthermore, the
phase diagram in Fig.~\ref{fig:phase_diagram} indicates how upon
raising the number of \addAK{self-peptides} there is a transition from
preference for strong amino-acids ($\beta>0$, positive selection dominant)
to weak amino-acids ($\beta<0$, negative selection dominant), which may be 
feasibly tested in future experiments, along the lines in Ref.~\cite{huseby06}.
}

This work was supported by National Institutes of Health (NIH) Grant
1-PO1-AI071195-01 and a NIH Director's Pioneer award (to A.K.C.).

\bibliography{EVD-PRL}
\end{document}